\documentclass[9pt,twocolumn,twoside]{osajnl}

\journal{ol} 

\setboolean{shortarticle}{true}

\title{Effective pair-interaction of phase singularities in random waves}

\author[1,2]{L. De Angelis}
\author[1,*]{L. Kuipers}

\affil[1]{Kavli Institute of Nanoscience, Delft University of Technology, 2600 GA, Delft, The Netherlands}
\affil[2]{The Netherlands Institute for Neuroscience, Institute of the Royal Netherlands Academy of Arts and Sciences (KNAW), Amsterdam, The Netherlands}

\affil[*]{Corresponding author: l.kuipers@tudelft.nl}

\begin{abstract}
In two-dimensional random waves, phase singularities are point-like dislocations with a behavior reminiscent of interacting particles. This -- qualitative -- consideration, stems from the spatial arrangement of these entities, which finds its hallmark in a pair correlation reminiscent of liquid-like system. Starting from their pair correlation function, we derive an effective pair-interaction for phase singularities in random waves by using a reverse Monte Carlo method. This study initiates a new approach for the treatment of singularities in random waves and can be generalized to topological defects in any system.
\end{abstract}

\setboolean{displaycopyright}{true}

\begin{document}

\maketitle

The phase singularities arising from the interference of monochromatic waves \cite{Nye1974} do not arrange in a completely random manner, even when the interfering waves are random \cite{Jain2017}.
Instead, the singularities' spatial distribution actually exhibits the hallmarks of a liquid-like system \cite{Berry2000}.
This behaviour of the phase singularities can be most clearly observed in the pair correlation function \cite{Berry2000,Saichev2001,Barth2002,Berggren2002,Kim2003,Kim2005,Hoehmann2009,DeAngelis2016,DeAngelis2018PRX,DeAngelis2019,vanGogh2020}, one of the most used tools to describe the spatial arrangement of discrete systems of any kind \cite{Barker1976,Hansen1990,Frenkel2001}.
Additionally, these singularities have a topological charge, also hinting at analogies with interacting particles, with mechanisms like same-charge repulsion \cite{Shvartsman1994,Freund1994Sign} and topological screening \cite{Freund1994,Dennis2003,VanTiggelen2006,DeAngelis2018}. 
However, we cannot directly measure the interaction of two singularities, and comparing them to interacting particles currently does not go beyond the superficial and the use of metaphors.

In this letter, we approach the determination of an effective interaction among phase singularities in a quantitative way.
We use a reverse-engineering approach \cite{Lyubartsev1995} to compute an effective pair-interaction for phase singularities in scalar random waves.
Starting with an initial guess for the pair-interaction, we use an iterative approach to amend the interaction potential until the potential generates an equilibrium configuration consistent with the input pair correlation.
We perform this procedure in the canonical ensemble, i.e., for a fixed number of singularities in a given observation area. 
With this choice, we are able to converge to a stable solution for the effective interaction among pairs of singularities, also taking into account their topological charge. 
Performing a final Monte Carlo simulation, using the solution for the effective interaction as an input, we are able to reproduce the theoretical pair correlation of phase singularities in random waves with excellent consistency. The method that we show here is general, and could also be applied to other systems than phase singularities in random waves.

\begin{figure}[htbp]
	\centering
	\includegraphics[width=\linewidth]{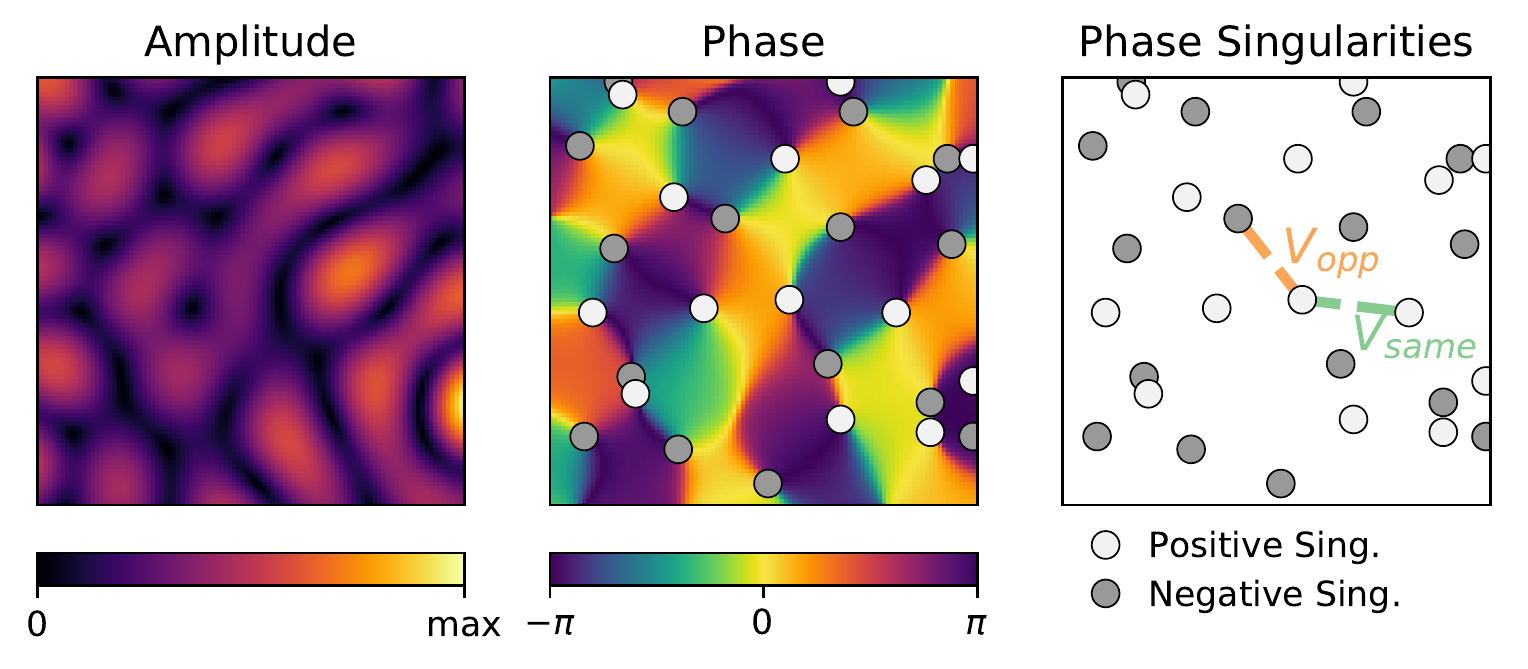}
	\caption{Amplitude, phase and singularities for a  calculated
		random wave field. The schematics in the third panel illustrates the
		effective pair interaction among singularities with the same ($V_{same}$) or
		opposite ($V_{opp}$) topological charge.}
	\label{fig:Veff_schem}
\end{figure}

Figure \ref{fig:Veff_schem} shows schematically how we can map a random wave field -- with its amplitude (left panel) and phase (middle panel) -- onto a distribution of interacting particles through a set of discrete points, which are its phase singularities (right panel).
Of course, in order for this conceptual step to be meaningful, the correct interaction needs to be provided.
We now retrieve such a fictive -- yet meaningful -- interaction by exploiting the knowledge on the pair correlation of phase singularities in random waves \cite{Berry2000, DeAngelis2016}. 
In particular, we will consider the model case of phase singularities in isotropic scalar waves, for which the pair correlation is known analitically \cite{Berry2000}.

Given the pair correlation function $g(r)$ of an ensemble of particles at their thermodynamic equilibrium, one can use various approximations to guess the underlying interaction.
In a mean-field approximation for instance \cite{Frenkel2001}, the interaction would simply be given by
\begin{equation}\label{e:mfield}
V^\mathrm{mf}(r) = -kT \log\,[g(r)].
\end{equation}
We hereby note that as in our system we do not have access to a parameter such as the temperature $T$, all the energies and interaction potentials that we compute will be expressed in units of $kT$, i.e., $kT=1$.
Although mean-field approximation can represent a good first step for a qualitative idea of the interaction in a system, it is a quite
coarse approximation. 
Using the mean-field expression for the pair interaction to simulate the evolution of a system in a Monte Carlo simulation does typically not reproduce the expected equilibrium state, which should be consistent with the initial pair correlation $g(r)$ \cite{Lyubartsev1995}.

One interesting approach to obtain the actual pair interaction $V(r)$ starting from a known pair correlation $g^{\mathrm{th}}(r)$, is to reverse engineer the problem of the Monte Carlo simulation. 
This method was already illustrated by Lyubartsev and Laaksonen in the context of ionic liquids physics \cite{Lyubartsev1995}. 
The procedure consists of three main steps. 
As a first step, one needs to start a preliminary Monte Carlo simulation with a guessed interaction $V^{(0)}(r)$. 
Subsequently, when the simulated system reaches its final equilibrium configuration, one can measure its pair correlation function $\langle g^{(0)}\rangle (r)$.
This pair correlation function can then be compared to the known $g^\mathrm{th}(r)$.
Based on the deviation $\Delta \langle g^{(0)}\rangle$ between the measured and expected pair correlation, the interaction potential $V^{(0)}(r)$ is suitably corrected, to obtain $V^{(1)}(r) = V^{(0)}(r) + \Delta V^{(0)}(r)$.
By iterating this process one can then calculate
\begin{equation}
V^{(j+1)}(r) = V^{(j)}(r) + \Delta V^{(j)}(r),
\end{equation}
until $\Delta V^{(j)}(r)$ becomes negligible, and convergence is reached.


The key step in the potential retrieval is to determine the right correction $\Delta V^{(j)}(r)$.
This correction was elegantly demonstrated to depend on the covariance of the measured pair correlation function~\cite{Lyubartsev1995}.
Explicitly, it is convenient to introduce the function $S(r)~=~2\pi r g(r)$ (valid for the 2D case), and expand it in its Taylor series in $\Delta V^{(j)}(r)$:
\begin{equation}
\Delta \langle S^{(j)}\rangle (r_\alpha)
= \sum_{r_\beta} \frac{\partial \langle S^{(j)} \rangle (r_\alpha)}{\partial V^{(j)}(r_\beta)}
\Delta V^{(j)}(r_\beta) + O(\Delta V^2).
\label{e:Sjtaylor}
\end{equation}
If we omit the terms of order $O(\Delta V^2)$, \eqref{e:Sjtaylor} is a system of linear equations in $\Delta V^{(j)}(r_\beta)$.
The problem is then translated into finding the coefficients of this system of linear equations:
\begin{equation}
    A^\alpha_\beta = \frac{\partial \langle S^{(j)}\rangle(r_\alpha)}{\partial V^{(j)}(r_\beta)}.
\end{equation}
Interestingly, by explictly writing the ensamble average $\langle\cdot\rangle$, one finds that these parameters depend on the covariance of $S$:
\begin{align}
\begin{aligned}
A^\alpha_\beta & = & 
\frac{\partial (r_\alpha)}{\partial V^{(j)}(r_\beta)} \left[ \frac{
\int dq\, S^{(j)}_q(r_\alpha)\, \exp \left[-\sum_{r_\gamma} V^{(j)}(r_\gamma) S^{(j)}_q(r_\gamma)\right]}{
\int dq\, \exp \left[-\sum_{r_\gamma} V^{(j)}(r_\gamma) S^{(j)}_q(r_\gamma)\right]}
\right]\\ \\
 & = &
- \left[ \langle S^{(j)}(r_\alpha)\, S^{(j)}(r_\beta) \rangle -
\langle S^{(j)}(r_\alpha)\rangle\,\langle S^{(j)}(r_\beta) \rangle\right].
\end{aligned}
\end{align}
Knowing the parameters $A^\alpha_\beta$, solving the system of linear
equations of \eqref{e:Sjtaylor} yields the corrections $\Delta V^{(j)}(r)$.

This algorithm was validated for systems of particles in three dimensions, both with and without the presence of a charge~\cite{Lyubartsev1995}.
When charge is present, it is convenient to split the problem in two subproblems, one for with the same charge, with pair correlation $g_\mathrm{same}(r)$ and pair interaction $V_\mathrm{same}(r)$, and one for particles with opposite charge, with $g_\mathrm{opp}(r)$ and $V_\mathrm{opp}(r)$. 
Since topological charge is relevant for the spatial distribution of phase singularities in random waves, and it is known that $g_\mathrm{same}(r) \neq g_\mathrm{opp}(r)$ \cite{DeAngelis2016}, we here make this distinction.


With the recipe for the iterative procedure, we use the theoretical pair correlation function of phase singularities in random waves \cite{Berry2000} to retrieve their effective interaction.
We perform the retrieval considering in parallel both cases of same and opposite topological charge interaction. 
Typically the mean-field potential is a good choice as a starting guess of the interaction potential $V^{(0)}(r)$ \cite{Lyubartsev1995}.
Including the topological charge, there would be two mean-field potentials:
\begin{equation}\label{e:chargemfield}
V^\mathrm{mf}_\mathrm{same}(r)  =  -\log\,[g_\mathrm{same}(r)],\quad
V^\mathrm{mf}_\mathrm{opp}(r)  =  -\log\,[g_\mathrm{opp}(r)],
\end{equation}
for same and oppositely charged pairs, respectively.
For the case of phase singularities, this choice of using the mean-field approximation does not lead to stable Monte Carlo simulations in a finite simulation box. This is due to the long-range correlations in $g_\mathrm{same}(r)$ and $g_\mathrm{opp}(r)$, which result in long-range oscillations in the corresponding mean-field potentials. Such oscillations make it more energetically favorable for singularities to be at a fixed distance and as a result the corresponding Monte Carlo simulation tend to form an ordered pattern with singularities. This results in a $g(r)$ too far from the expected $g^{th}(r)$ to be handled by the retrieval algorithm, based on perturbation theory. We find that truncating these oscillations, e.g. setting $V(r > r_c) = 0$, solves this convergence issue. We checked that within a reasonable range that allows us to retain the interaction below the typical inter-singularity distance ($r_c > 0.5$) while cutting the long-range oscillations ($r_c < 3$), the specific choice of $r_c$ does not impact the final result.

In our study we do not allow for creation and annihilation of pairs of phase singularities and therefore do not allow for fluctuations of their average density. Further developments on this topic could include creation/annihilation events, which do not necessarily result in density fluctuations \cite{Kuhl2007}.
Practically, we set the density of singularities to be equal to its theoretical average value: $d = \pi/\lambda^2$, where $\lambda$ is the wavelength of the wave field \cite{Berry2000}, which we set to unity for convenience.

\begin{figure*}[thb]
	\centering
	\includegraphics[width=\linewidth]{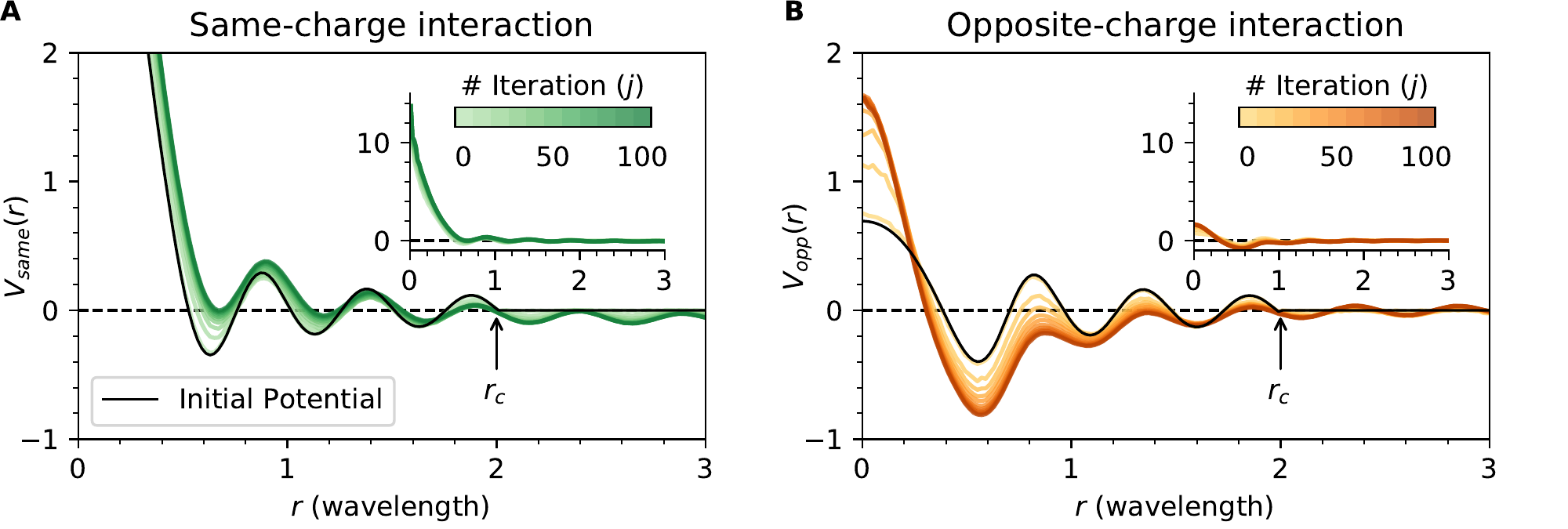}
	\caption{Overview of the retrieval of the effective potential between
		same-charge (A) and opposite-charge (B) singularities. The main plots display a close look on the main features of the two interaction potentials, whereas the insets depict their entire behavior. The color
		palettes in both panels represent the iteration number, from
		light colors ($j= 0$) to dark ones (up to $j = 100$). The black solid lines indicate the initial potential, and the black arrows highlights the point at which it is truncated compared to the mean field potential. The data for each iteration is available at \cite{Data}.}
	\label{fig:Veff}
\end{figure*}

At every step $j$ of the potential retrieval procedure, we perform a full Monte Carlo simulation of the system, based on the interaction potentials $V_\mathrm{same}^{(j)}(r)$ and $V_\mathrm{opp}^{(j)}(r)$.
Given these pair interactions of the system, we run a Monte Carlo simulation to compute its final state at equilibrium \cite{Frenkel2001}. 
The Monte Carlo simulation is carried in a square region of side $L = 10$ ($\lambda = 1$) with periodic boundary conditions. We note that we performed preliminary simulations with different sizes of the simulation box ($L = 5,\,10,\,20$) and verified that this size does not affect the final results.
The simulation box contains a number of particles $N=\rho / L^2 = 314$, where $\rho$ is the expected average density of singularities in isotropic random waves \cite{Berry2000}. To initialize the simulation, such particles are arranged in a square lattice with a pitch $a \approx L/\sqrt{N}$, perturbed with additive random displacements $\Delta x$ and $\Delta y$ uniformly distributed in $[-a/4,a/4]$.
Subsequently, the simulation is carried by proposing at each Monte Carlo step 
a displacement $\Delta r$ for the particle located at $r_i$. 
We compute the difference in energy $\Delta E$ corresponding to this change
\begin{equation}
\Delta E = \sum_{k\;\in\;\mathrm{system}} \left[V^{(j)}(r_i + \Delta r - r_k) - V^{(j)}(r_i - r_k) \right] ,
\end{equation}
and accept the change with a probability
\begin{equation}
p_{\mathrm{acc}} = \min \left[ 1,\,\exp\left(-\frac{\Delta E}{kT}\right) \right].
\end{equation}
This is a notorious method, known as the Metropolis algorithm~\cite{Frenkel2001}.
The total energy of the system is monitored after each Monte Carlo step, and it is used to determine if the equilibrium configuration has been reached. 
This we estimate as the moment in which the total energy does not exhibit significant changes rather than statistical fluctuations.
For each Monte Carlo simulation we perform $10^7$ Monte Carlo steps, and typically we observe that the equilibrium configuration is reached after approximately $10^5$ Monte Carlo steps (not shown).

Figure \ref{fig:Veff} presents an overview of the outcome of our iterative procedure for the retrieval of the effective interaction among phase singularities in random waves. 
The green and orange lines are the resulting potentials after each iteration $j$, changing from light to dark colors as $j$ is increased (only one in every ten iteration is shown).
On the left we depict the interaction potential among same-charge singularities, while on the right the interaction among oppositely charged ones is displayed. 
From both plots, we can observe how after a few tens of iterations the algorithm starts converging to a stable solution.
Interestingly, the oscillations in $V(r)$ that we forcefully suppressed in the starting potential $V^{(0)}$, reappeared as a result of the potential retrieval. 
In Fig.~\ref{fig:Veff} we can see how for the lighter colors (first iterations) $V(r > 2) = 0$, due to the initial truncation.
The differences between the mean-field potential and the final $V(r)$ are interesting, and are actually crucial for the retrieved potential to reproduce the behavior of singularities in random waves. 
One clear difference in both $V_\mathrm{same}(r)$ and $V_\mathrm{opp}(r)$ is that the average value around which the potential oscillates as a function of $r$ is not fixed at 0 as in the mean-field approximation. 
This average value decreases as $r$ is increased for $V_\mathrm{same}(r)$, whereas it increases with $r$ for $V_\mathrm{opp}(r)$. 
This is especially clear from the first dip in $V_\mathrm{same}(r)$, which has been significantly lifted by the retrieval procedure.
One might notice that as $r$ increases, $V_\mathrm{same}(r)$ approaches a negative value. We stress that this value should not be attributed any particular meaning, as interaction potentials - as opposed to correlation functions - are defined up to a constant.
With respect to $V_\mathrm{opp}(r)$, the additional clear difference implemented by the retrieval algorithm is the higher contrast between the first dip and its value at $r = 0$.

In both cases of $V_\mathrm{same}(r)$ and $V_\mathrm{opp}(r)$, most of the changes happen in proximity of the first dip in the potential, which is indeed the most relevant as it dominates the nearest-neighbors interaction. 
It is not immediately obvious how to interpret the oscillations that follow the first dip, and it might be interesting to further investigate on this feature in the future. 
In this context it is important to point out that these oscillations can also be observed in the effective-interaction description of an ionic-liquid in a polarizable solvent \cite{Lyubartsev1995}.

The prime test to perform in order to verify the validity of the interaction potential that we computed in this section, is to perform a final Monte Carlo simulation based on this interaction, and to compare the equilibrium configuration of this simulation with the theoretical spatial distribution of phase singularities in random waves.
\begin{figure*}[bthp]
	\centering
	\includegraphics[width=\textwidth]{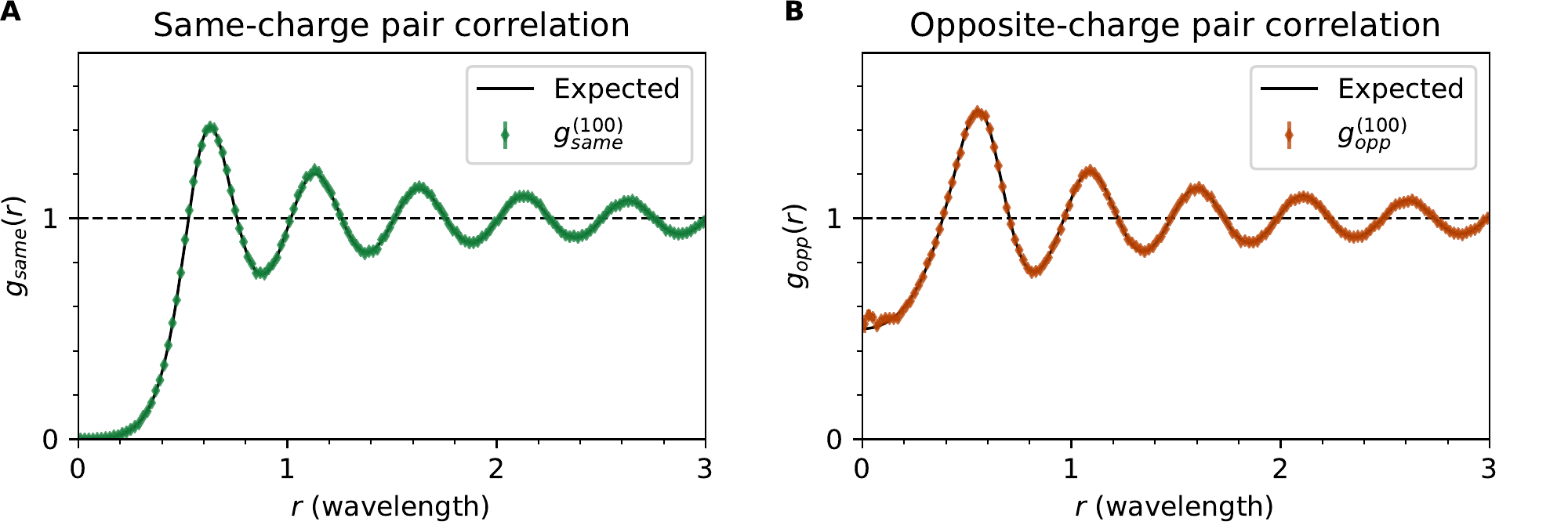}
	\caption{Pair correlation function measured at the iteration $j=100$ (data points), in
		direct comparison with the expected functions (solid lines). On the left (A)
		the correlation for same-charge singularities and on the right (B) the correlation for opposite-charge. The data for each iteration is available at \cite{Data}.}
	\label{fig:g_conv}
\end{figure*}
Figure \ref{fig:g_conv} directly compares the pair distribution functions computed on the outcome of the Monte Carlo simulation with the theoretical curves $g^{th}(r)$, for both same-charge and opposite-charge singularities. 
In both cases we observe a very good agreement, certifying that the retrieved potential is the one that can reproduce the spatial correlation of phase singularities in random waves.

\begin{figure}[h!tbp]
	\centering
	\includegraphics[width=\linewidth]{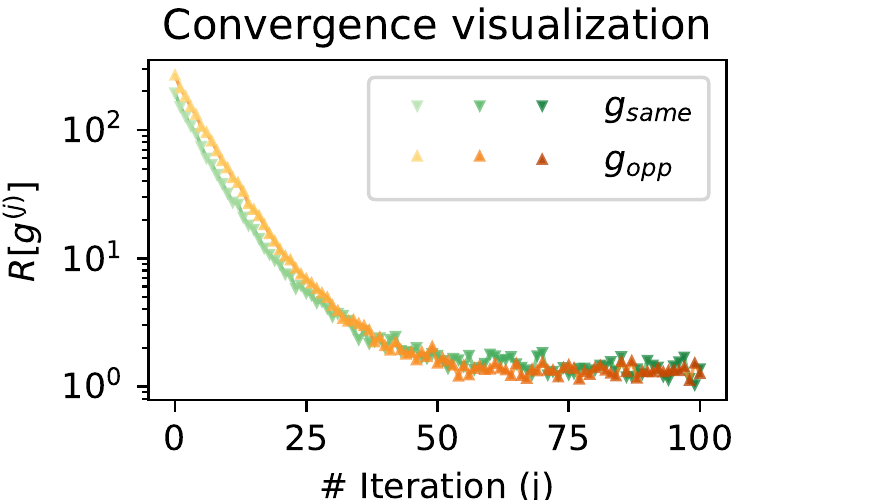}
	\caption{Visualization of the convergence for the iterative procedure. The plot shows the average residual
		$R[ g^{(j)}]$ between measured and expected pair correlation function
		at every iteration $j$, for both same- and opposite-sign singularities.}
	\label{fig:Veff_conv}
\end{figure}

We use the comparison between the measured pair correlations and the theoretical ones also as a way to quantify the convergence of our iterative algorithm.
Precisely, we consider the total residual $R[g^{(j)}]$ of the observed pair correlations $g^{(j)}(r)$ with respect to their expectation values $g^{\mathrm{th}}(r)$:
\begin{equation}
R[ g^{(j)}]
		= \frac{1}{N_\alpha} \sum_{r_\alpha} \frac{|g^{(j)}(r_\alpha) - g^\mathrm{th}(r_\alpha)|^2}{
									\sigma[g^{(j)}(r_\alpha)]^2},
\end{equation}
where $N_\alpha$ is the available number of discrete radii $r_\alpha$, and $\sigma[g^{(j)}(r_\alpha)]$ the uncertainty on each observed $g^{(j)}(r_\alpha)$, estimated with the standard deviation on independent sampling of the equilibrium configuration of the Monte Carlo simulation. 
Figure~\ref{fig:Veff_conv} shows the total residual $R$ for each iteration $j$. 
As we can see, a reasonable convergence is already obtained for $j\approx 60$.

We demonstrated how phase singularities in random waves can be thought of as of interacting particles, not only qualitatively, but also quantitatively. 
The results of this letter are shown for the prime example of singularities in isotropic random waves, a well known system already for a few decades \cite{Berry2000}. 
It is important to emphasize that the method shown here is general, as it only requires a knowledge of the pair correlation functions $g(r)$.
It is indeed known that singularities in any generic field can behave differently from those in isotropic random waves. This is the case for singularities in the vector components of a random light field, where the distribution of singularities is anisotropic and therefore a knowledge of the angle dependent pair correlation function is needed \cite{DeAngelis2016}.
Similarly, for polarization singularities in 2D random light, where singularities with different topological charge exhibit different spatial correlations \cite{DeAngelis2018PRX,DeAngelis2019}. All of the specific  properties can now be mapped into interaction potentials, possibly offering further insights on the role of singularities in wave fields.

We thank F. Alpeggiani
and B. Mulder and M.A. van Gogh for critical reading of the manuscript and useful discussions.
This work is part of the research
program of the Netherlands Organization for Scientific Research (NWO).
The authors acknowledge
funding from FP7 Ideas: European Research Council (ERC Advanced Grant
No. 340438-CONSTANS).

\bigskip \noindent
{\bf \large Disclosures.} The authors declare no conflicts of interest.

\bigskip \noindent
{\bf \large Data availability.} Data and code underlying the results presented in this paper are available in Ref.~\cite{Data}.

\bibliography{references}
\nocite{*}


\end{document}